\documentclass[10pt,twocolumn]{article}

\usepackage[utf8]{inputenc}
\usepackage[T1]{fontenc}
\usepackage{microtype}
\usepackage{times}
\usepackage[margin=2cm]{geometry}

\usepackage{amsmath,amssymb}

\usepackage{graphicx}
\graphicspath{{figures/}}
\usepackage[dvipsnames]{xcolor}

\usepackage{booktabs}
\usepackage{multirow}
\usepackage{array}

\usepackage[colorlinks,citecolor=MidnightBlue,linkcolor=BrickRed,urlcolor=MidnightBlue]{hyperref}

\usepackage{enumitem}
\setlist{nosep,leftmargin=*}

\usepackage[font=small,labelfont=bf]{caption}

\usepackage{algorithm}
\usepackage{algorithmic}

\usepackage{xspace}
\newcommand{\ie}{\emph{i.e.}\xspace}
\newcommand{\eg}{\emph{e.g.}\xspace}

\title{Cluster-First Labelling: An Automated Pipeline for\\
       Segmentation and Morphological Clustering\\
       in Histology Whole Slide Images}

\author{
  \begin{tabular}{c}
    Muhammad Haseeb Ahmad \\
    \small Department of Engineering Science \\
    \small \texttt{haseeb.ahmad@eng.ox.ac.uk} \\[6pt]
    Damion Young \\
    \small Medical Sciences Division \\
    \small \texttt{damion.young@medsci.ox.ac.uk}
  \end{tabular}
  \and
  \begin{tabular}{c}
    Sharmila Rajendran \\
    \small Department of Physiology, Anatomy and Genetics \\
    \small \texttt{sharmila.rajendran@dpag.ox.ac.uk} \\[6pt]
    Jon Mason \\
    \small Medical Sciences Division \\
    \small \texttt{jon.mason@medsci.ox.ac.uk}
  \end{tabular}
} 

\date{\small University of Oxford, United Kingdom}

\begin{document}
\maketitle

\begin{abstract}
Labelling tissue components in histology whole slide images (WSIs) is
prohibitively labour-intensive: a single slide may contain tens of
thousands of structures---cells, nuclei, and other morphologically
distinct objects---each requiring manual boundary delineation and
classification.
We present a cloud-native, end-to-end pipeline that automates this
process through a \emph{cluster-first} paradigm.
Our system tiles WSIs, filters out tiles deemed unlikely to contain
valuable information, segments tissue components with Cellpose-SAM
(including cells, nuclei, and other morphologically similar
structures), extracts neural embeddings via a pretrained ResNet-50,
reduces dimensionality with UMAP, and groups morphologically similar
objects using DBSCAN clustering.
Under this paradigm, a human annotator labels representative clusters
rather than individual objects, reducing annotation effort by orders of
magnitude.
We evaluate the pipeline on 3{,}696 tissue components across 13 diverse
tissue types from three species (human, rat, rabbit), measuring how
well unsupervised clusters align with independent human labels via
per-tile Hungarian-algorithm matching.
Our system achieves a weighted cluster--label alignment accuracy of
\textbf{96.8\%}, with 7 of 13 tissue types reaching perfect agreement.
The pipeline, a companion labelling web application, and all evaluation
code are released as open-source software.
\end{abstract}

\section{Introduction}
\label{sec:intro}

Whole slide imaging has transformed histopathology by digitising glass
slides at high resolution, enabling computational analysis at
scale~\cite{pantanowitz2011wsi,farahani2015wsi}.
A critical downstream task is \emph{cell-level annotation}: identifying
individual cells, delineating their boundaries, and assigning
morphological or functional labels.
These annotations provide a valuable educational resource for medical
students in our setting.

\paragraph{Terminology.}
Cell-boundary detection models such as Cellpose segment any
morphologically distinct structure that resembles a cell, including
individual cells, nuclei, clusters of tightly packed cells, and
other tissue components.
Filtering these heterogeneous detections at the point of segmentation
is impractical without domain-specific heuristics.
Throughout this paper we use \emph{cell} as convenient shorthand for
any such segmented object; where the distinction matters we write
\emph{tissue component} or qualify explicitly.

Manual annotation remains the dominant approach, yet it is
prohibitively expensive.
A single WSI scanned at $40\times$ magnification can span
$100{,}000 \times 100{,}000$ pixels and contain tens of thousands of
cells.
Tracing boundaries and classifying each cell individually can take
days of expert time per slide.

\paragraph{Cluster-first labelling.}
We propose an alternative paradigm: rather than labelling cells one at
a time, our pipeline \emph{first} segments all cells automatically,
\emph{then} groups morphologically similar cells into clusters.
A human annotator need only review and label each cluster once; the
label propagates to every member of that cluster.
Because morphologically distinct tissue components---such as 
nuclei or tightly packed cell groups---are
naturally separated into their own clusters, an annotator can label or
discard entire categories at the cluster level rather than inspecting
individual objects.
For example, if a slide contains 15{,}000 segmented objects grouped
into 25 clusters, the annotator reviews 25 representative groups
instead of 15{,}000 individual objects---a reduction of roughly
$600\times$.

\paragraph{Contributions.}
\begin{enumerate}
  \item An end-to-end, cloud-native pipeline that takes raw WSI files
        and produces per-cell cluster assignments, requiring no manual
        intervention (Section~\ref{sec:methods}).
  \item A scalable Azure ML implementation supporting multi-node
        parallelism with per-slide granularity
        (Section~\ref{sec:scalability}).
  \item An open-source web application for human validation that
        computes per-tile Hungarian-aligned accuracy between
        unsupervised clusters and human labels
        (Section~\ref{sec:eval_framework}).
  \item Empirical evaluation on 3{,}696 cells from 13 tissue types
        demonstrating 96.8\% cluster--label alignment accuracy
        (Section~\ref{sec:experiments}).
\end{enumerate}

\section{Related Work}
\label{sec:related}

\paragraph{Cell segmentation.}
Classical approaches relying on thresholding struggle in complex 
environments, and watershed algorithm struggles with over segmentation.
Deep-learning methods such as U-Net~\cite{ronneberger2015unet},
StarDist~\cite{schmidt2018stardist}, and
Hover-Net~\cite{graham2019hovernet} have significantly improved
accuracy.
Cellpose~\cite{stringer2021cellpose} introduced a gradient-flow
representation that generalises across cell types without
retraining.
Its latest variant, Cellpose-SAM~\cite{pachitariu2025cellposesam},
integrates a Segment Anything backbone to achieve state-of-the-art
generalisation, which we adopt in our pipeline.

\paragraph{Computational pathology pipelines.}
CLAM~\cite{lu2021clam} pioneered attention-based weakly supervised
learning on WSIs but operates at the slide level rather than the cell
level.
QuPath~\cite{bankhead2017qupath} provides interactive cell detection
and measurement but requires substantial manual configuration per
tissue type.
Foundation models for pathology such as
UNI~\cite{chen2024pathology} offer domain-specific embeddings and
are publicly available, albeit behind gated access with restrictive
non-commercial licences.
Our work is complementary: we combine off-the-shelf, permissively
licensed components (Cellpose-SAM, ImageNet ResNet-50) into a fully
automated pipeline with an explicit clustering step that enables the
cluster-first annotation paradigm.
Because the downstream task is unsupervised clustering of relative
morphological similarity rather than fine-grained classification, a
general-purpose backbone such as ResNet-50 provides sufficient
discriminative power while ensuring unrestricted reproducibility.

\paragraph{Unsupervised cell grouping.}
Extracting meaningful representations of cellular morphology without
task-specific labels is a persistent challenge in digital
pathology. Unsupervised representation learning has recently 
shown strong potential for categorizing complex tissue 
phenotypes~\cite{srinidhi2021deep}. However, many studies 
focus on tile-level analysis rather than single-cell 
morphological grouping, or they evaluate clustering on 
pre-cropped cell datasets rather than bridging the gap 
to full WSIs~\cite{doron2023unbiased}. By chaining state-of-the-art segmentation, feature 
extraction, and clustering into a unified pipeline, we move beyond 
disjointed tooling. Furthermore, our system pairs 
unsupervised cell characterisation with a purpose-built human-in-the-loop 
validation interface that directly consumes pipeline outputs, 
addressing the practical difficulties of 
verifying morphological cell clusters.

\section{System Architecture}
\label{sec:methods}

\begin{figure}[t]
  \centering
  \includegraphics[width=\linewidth]{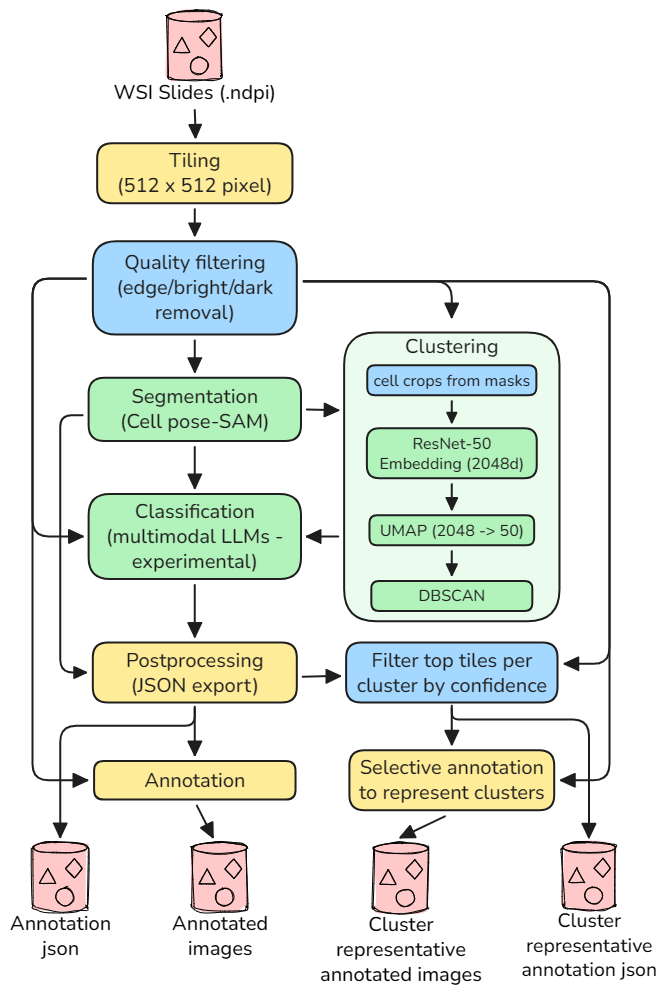}
  \caption{%
    Pipeline architecture.  WSI slides (\texttt{.ndpi}) are tiled into
    $512 \times 512$ patches, quality-filtered, segmented with
    Cellpose-SAM, embedded with ResNet-50, reduced via UMAP, and
    clustered with DBSCAN.  Optional downstream stages annotate images,
    extract representative tiles per cluster, and invoke a multimodal
    LLM for experimental classification.
  }
  \label{fig:pipeline}
\end{figure}

Figure~\ref{fig:pipeline} shows the end-to-end pipeline.
Each stage is implemented as an independent, containerised component
orchestrated by Azure Machine Learning~\cite{azureml2023}.

\subsection{WSI Tiling and Quality Filtering}
\label{sec:tiling}

Raw WSI files are read via
OpenSlide~\cite{goode2013openslide} and partitioned into
non-overlapping $512 \times 512$-pixel tiles.
The pipeline supports multiple magnification levels via downsampling or
upsampling factors relative to the native resolution, and optional
uniform sub-sampling to limit the tile count per magnification.

An optional quality-filtering stage removes uninformative tiles using
six image-quality metrics: edge density
(Canny~\cite{bradski2000opencv}), bright-pixel ratio,
dark-pixel ratio, intensity standard deviation, Laplacian
variance~\cite{bradski2000opencv} (focus quality), and cross-channel
colour variance.
Tiles failing any threshold are discarded before segmentation, reducing
computation on background and out-of-focus regions.

\subsection{Cell Segmentation}
\label{sec:segmentation}

We use Cellpose-SAM~\cite{pachitariu2025cellposesam} (\texttt{cpsam}
model) for boundary detection of cell-like structures.
Cellpose-SAM combines the Cellpose gradient-flow formulation with a
Segment Anything encoder, achieving robust generalisation across tissue
types without fine-tuning.
The model detects a variety of objects resembling cells, including 
individual cells but also nuclei, tightly packed cell groups, and other 
morphologically distinct tissue components depending on visual context.
Discriminating these categories at the segmentation stage is infeasible
without tissue-specific filtering rules; instead, the downstream
clustering step naturally groups them into separate, visually coherent
clusters that can be labelled or discarded at the cluster level.

For each tile, the model produces an instance segmentation mask from
which we extract per-object bounding boxes and boundary polygons.
Key parameters---flow threshold ($0.4$), cell-probability threshold
($0.0$), and optional diameter estimation---are exposed as pipeline
arguments to support tissue-specific tuning.
GPU acceleration is used by default; tiles can be batched for higher
throughput on high-VRAM hardware.

\subsection{Neural Embedding and Clustering}
\label{sec:clustering}

\paragraph{Feature extraction.}
Each segmented cell is cropped from its parent tile using its bounding
box and passed through a ResNet-50~\cite{he2016resnet} backbone
pretrained on ImageNet~\cite{deng2009imagenet}.
We chose ResNet-50 for its well-established performance on visual
feature extraction and wide availability across deep-learning frameworks.
Additionally, because the downstream clustering operates on relative
morphological similarity rather than absolute feature quality---
the choice of backbone is unlikely to materially affect results.
We extract the 2{,}048-dimensional output of the penultimate
average-pooling layer as the cell's feature vector.

\paragraph{Dimensionality reduction.}
When enabled, UMAP~\cite{mcinnes2018umap} projects the 2{,}048-dimensional
embeddings to 50 dimensions while preserving local and global
morphological structure.
Default hyper-parameters are $k = 15$ neighbours,
$d_{\min} = 0.1$, Euclidean metric.

\paragraph{Clustering.}
DBSCAN~\cite{ester1996dbscan} groups cells into clusters without
requiring a predefined number of classes.
The neighbourhood radius $\varepsilon$ is estimated automatically via
the knee-point of the sorted $k$-nearest-neighbour distance curve; the
minimum core-point size is set to $\mathtt{min\_samples} = 5$.
When a CUDA-capable GPU is available, the RAPIDS
cuML~\cite{rapids2019} implementation is used for acceleration.
Cells that do not meet the density criteria are labelled as noise
(cluster $-1$).

\subsection{Scalability}
\label{sec:scalability}

The pipeline supports two execution modes selected by a single
parameter (\texttt{--max\_nodes}):
\begin{itemize}
  \item \textbf{Sequential} ($n = 1$): each stage runs as a single
        Azure ML \texttt{command()} job with linear logging, useful for
        debugging.
  \item \textbf{Parallel} ($n > 1$): stages use Azure ML
        \texttt{parallel\_run\_function()} with slide-level
        granularity, distributing slides across up to $n$ GPU nodes.
\end{itemize}
Both modes execute the same entry-point scripts; only the orchestration
wrapper differs.

\subsection{Human Evaluation Framework}
\label{sec:eval_framework}

\begin{figure}[t]
  \centering
  \includegraphics[width=\linewidth]{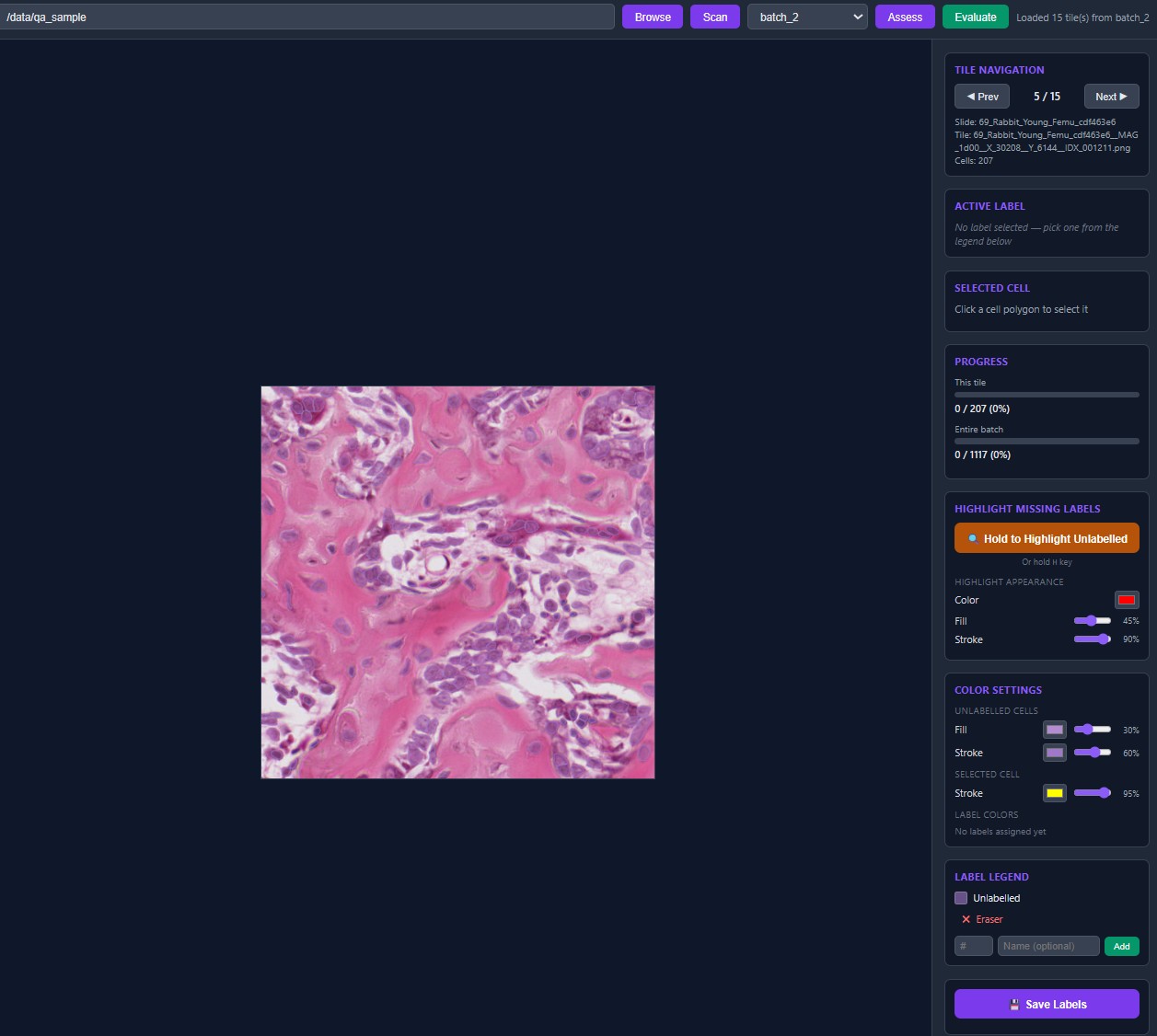}
  \caption{%
    Labelling application interface.  The central panel shows a
    $512 \times 512$ histology tile with interactive cell polygons.
    The sidebar provides tile navigation, label assignment controls,
    per-tile and batch progress, and visual customisation settings.
  }
  \label{fig:labelling_app}
\end{figure}

To measure how well unsupervised clusters align with human judgement,
we developed a companion web application
(Figure~\ref{fig:labelling_app}).
Built with FastAPI, the application:

\begin{enumerate}
  \item Displays sampled tiles with segmentation-polygon overlays.
  \item Lets the annotator define label classes and assign them to
        cells by clicking on polygons.
  \item Exports human labels as timestamped JSON files.
  \item Computes accuracy via per-tile Hungarian-algorithm
        matching~\cite{kuhn1955hungarian}, then aggregates tile-level
        results into an overall batch-level accuracy score.
\end{enumerate}

\paragraph{Hungarian-aligned accuracy.}
Because unsupervised cluster IDs are arbitrary, direct comparison with
human labels is meaningless.
For each tile, we build a contingency matrix $\mathbf{C} \in
\mathbb{Z}^{k \times k}$ where $k = \max(m, h)$, $m$ is the number of
model clusters and $h$ the number of human labels, with $C_{ij}$
counting cells assigned to model cluster~$i$ and human label~$j$
(entries involving padding rows or columns are zero).
The Hungarian algorithm~\cite{kuhn1955hungarian} finds the optimal
one-to-one mapping $\pi^*$ from model clusters to human labels that
maximises total agreement:
\begin{equation}
  \pi^* = \arg\max_{\pi} \sum_{i} C_{i,\,\pi(i)}
  \label{eq:hungarian}
\end{equation}
Accuracy is then the fraction of cells whose mapped model cluster
matches their human label.
Per-tile alignment is preferred over global alignment because cluster
IDs may vary across slides and tiles.

\section{Experiments}
\label{sec:experiments}

\subsection{Experimental Setup}
\label{sec:setup}

\paragraph{Data.}
We processed WSI files (\texttt{.ndpi} format) from 13 slides spanning
13 distinct tissue types and three species (Table~\ref{tab:per_slide}).
No slide-specific parameter tuning was performed; the same fixed
configuration was used for all tissue types.

\paragraph{Pipeline configuration.}
Tile quality filtering was enabled using the default thresholds for
all six metrics described in Section~\ref{sec:tiling}.
Segmentation used Cellpose-SAM (\texttt{cpsam}) with flow
threshold~$0.4$, cell-probability threshold~$0.0$, and automatic
diameter estimation.
Embeddings were extracted with a ResNet-50 pretrained on ImageNet
and L2-normalised prior to clustering.
UMAP reduced dimensionality from 2{,}048 to 50.
DBSCAN used $\varepsilon = 0.5$ and
$\mathtt{min\_samples} = 5$; clustering was performed independently
per slide rather than globally.
GPU acceleration was enabled for both segmentation (Cellpose) and
clustering (RAPIDS cuML).
The optional LLM classification stage was disabled
($\mathtt{classify\_per\_cluster} = 0$).
Post-processing aggregated segmentation and clustering outputs into
structured per-slide JSON files.
Annotated tile images were generated with cluster-ID-coloured polygon
overlays.
A representative-tile extraction step selected up to 10 cells per
cluster with no minimum confidence threshold
($\mathtt{confidence\_threshold} = 0.0$), and a second annotation pass
rendered polygon overlays on the filtered representative tiles.

\paragraph{Evaluation.}
Pipeline outputs were sampled into three evaluation batches using
slide-balanced random sampling to ensure every tissue type was
represented.
Each slide was allocated an equal cell budget
($\lceil 3{,}000 / 13 \rceil \approx 231$ cells) regardless of its
total cell count; tiles within each slide were randomly shuffled and
greedily selected until the budget was met.
The pooled selection was then shuffled globally and partitioned into
three batches of approximately 1{,}000 cells each.
A human annotator independently labelled all cells in each batch tile
using the labelling application, without access to cluster assignments.
Hungarian-aligned accuracy (Eq.~\ref{eq:hungarian}) was computed per
tile and aggregated per batch.

\subsection{Results}
\label{sec:results}

\paragraph{Batch-level results.}
Table~\ref{tab:batch} summarises the three evaluation batches.
The pipeline achieves consistently high cluster--label alignment, with
a weighted mean accuracy of \textbf{96.8\%} across 3{,}696 cells.

\begin{table}[t]
  \centering
  \caption{%
    Batch-level evaluation results.  Accuracy is the weighted
    Hungarian-aligned accuracy across all tiles in each batch.
  }
  \label{tab:batch}
  \small
  \begin{tabular}{@{}lrrrrr@{}}
    \toprule
    \textbf{Batch} & \textbf{Cells} & \textbf{Tiles}
      & \textbf{Human} & \textbf{Model}
      & \textbf{Acc.\ (\%)} \\
    & & & \textbf{clusters} & \textbf{clusters} & \\
    \midrule
    Batch 1 & 1{,}229 &  5 & 2 & 1 & 94.9 \\
    Batch 2 & 1{,}117 & 15 & 3 & 2 & 97.7 \\
    Batch 3 & 1{,}350 &  9 & 2 & 1 & 97.8 \\
    \midrule
    \textbf{Total} & \textbf{3{,}696} & \textbf{29}
      & --- & --- & \textbf{96.8} \\
    \bottomrule
  \end{tabular}
\end{table}

\paragraph{Per-tissue analysis.}
Table~\ref{tab:per_slide} breaks down accuracy by tissue type.
Seven of 13 tissue types achieve \emph{perfect} cluster--label
agreement (100\%).
Performance is highest on tissues with well-separated, homogeneous cell
populations (\eg lung, prostate, cervix) and lowest on compact bone
and skeletal muscle, which contain densely packed, morphologically
variable structures.

\begin{table}[t]
  \centering
  \caption{%
    Per-tissue-type evaluation.  Accuracy is the weighted
    Hungarian-aligned accuracy over all tiles from that slide, pooled
    across batches.
  }
  \label{tab:per_slide}
  \small
  \begin{tabular}{@{}llrrr@{}}
    \toprule
    \textbf{Tissue type} & \textbf{Species}
      & \textbf{Cells} & \textbf{Tiles} & \textbf{Acc.\ (\%)} \\
    \midrule
    Appendix        & Human  &   662 &  1 & 91.4 \\
    Ovary           & Human  &   587 &  2 & 99.7 \\
    Compact bone    & Human  &   106 & 11 & 84.0 \\
    Skeletal muscle & Human  &   219 &  3 & 84.0 \\
    Pancreas        & Human  &    28 &  1 &100.0 \\
    Kidney          & Rat    &   328 &  2 & 97.9 \\
    Prostate        & Human  &   315 &  1 &100.0 \\
    Cervix          & Human  &   240 &  2 &100.0 \\
    Femur           & Rabbit &   395 &  2 & 99.7 \\
    Lung            & Human  &   218 &  1 &100.0 \\
    Submandib.\ gland & Human &  178 &  1 &100.0 \\
    Seminal vesicle & Human  &   381 &  1 &100.0 \\
    Fallopian tube  & Human  &    39 &  1 &100.0 \\
    \midrule
    \textbf{All tissues} & ---
      & \textbf{3{,}696} & \textbf{29} & \textbf{96.8} \\
    \bottomrule
  \end{tabular}
\end{table}

\begin{figure}[t]
  \centering
  \includegraphics[width=\linewidth]{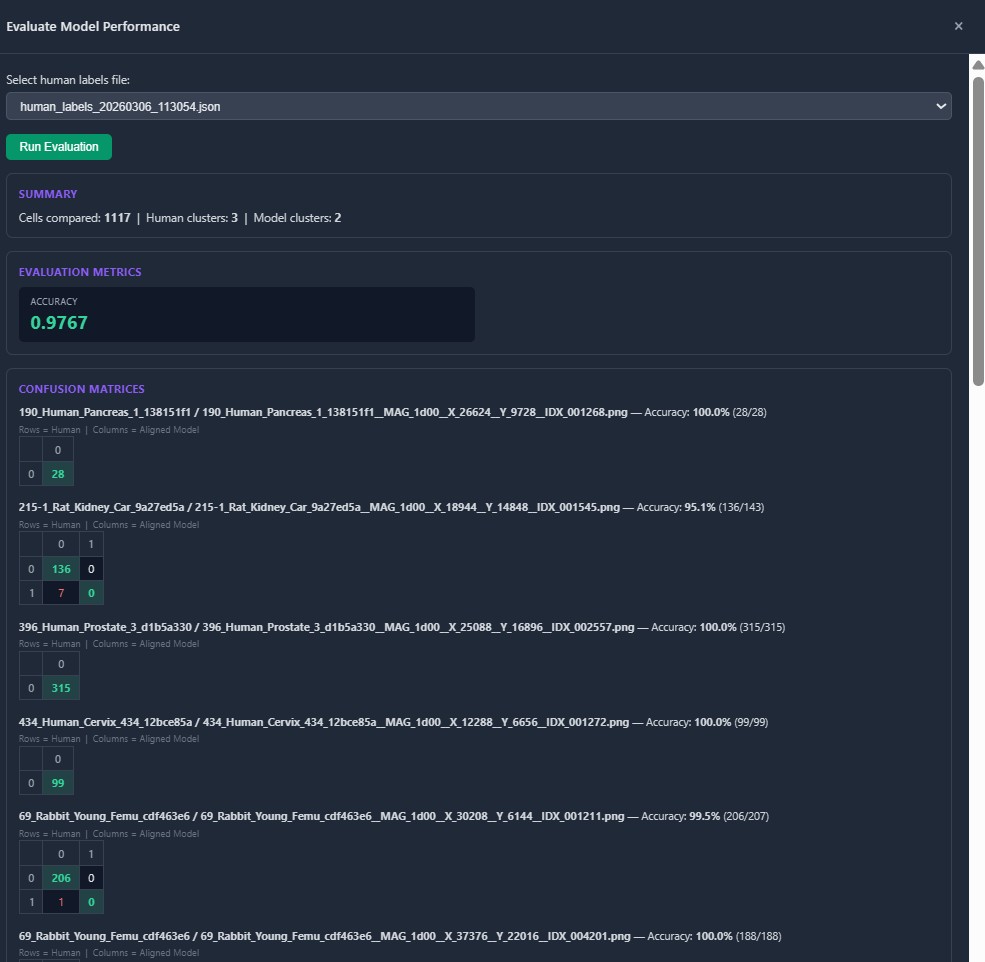}
  \caption{%
    Evaluation interface showing batch-level accuracy (97.67\%) and
    per-tile confusion matrices.  Rows represent human labels; columns
    represent Hungarian-aligned model clusters.  Green cells indicate
    correct assignments.
  }
  \label{fig:confusion}
\end{figure}

\paragraph{Representative confusion matrices.}
Figure~\ref{fig:confusion} shows per-tile confusion matrices from the
evaluation application.
In most tiles the matrix is diagonal or near-diagonal, confirming
strong cluster--label alignment.
Off-diagonal entries are concentrated in compact bone and skeletal
muscle tiles.

\begin{figure}[t]
  \centering
  \includegraphics[width=\linewidth]{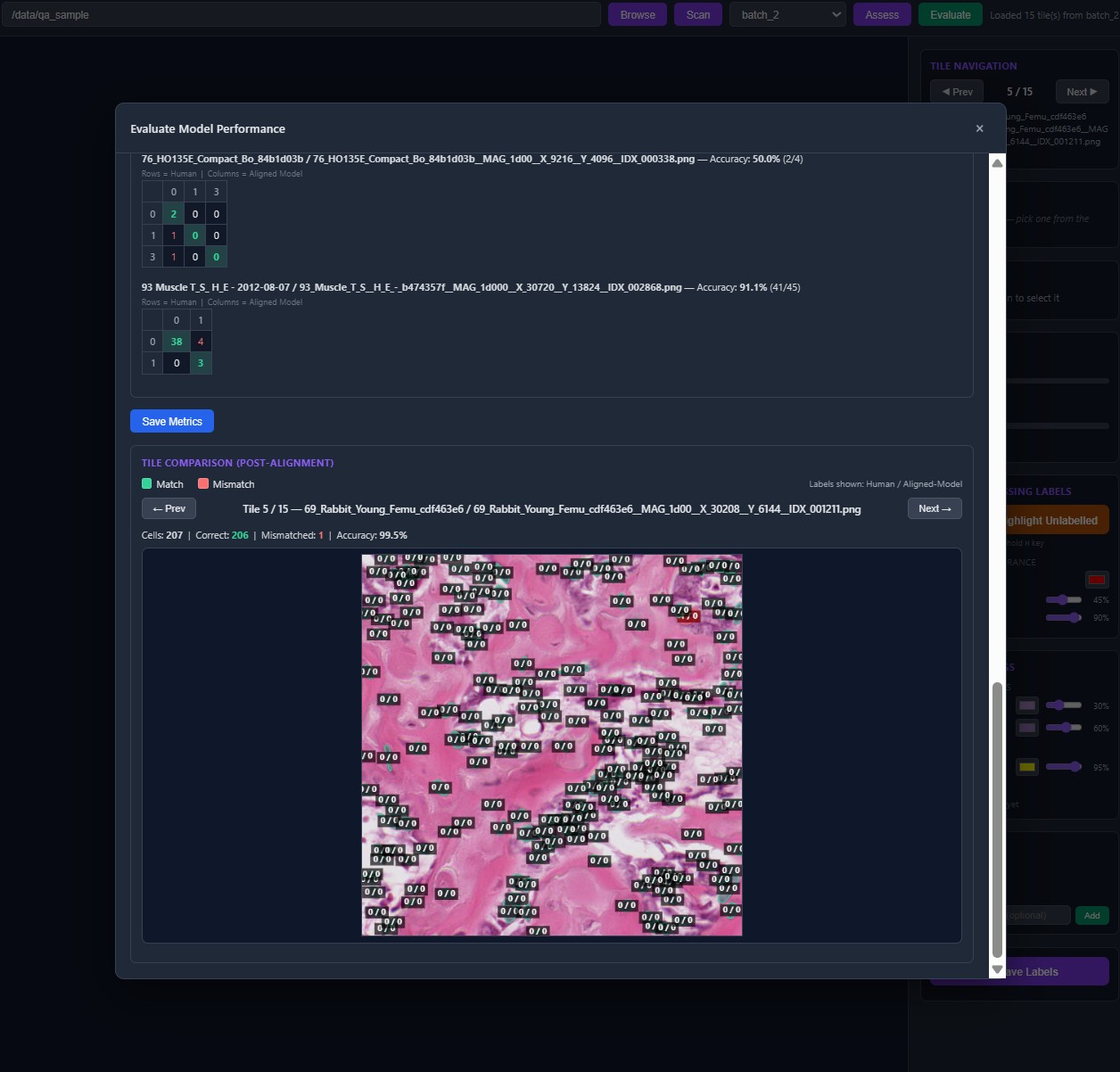}
  \caption{%
    Post-alignment tile comparison.  Cell polygons are colour-coded:
    green = human and aligned-model labels agree; red = disagree.
    Labels are shown as \emph{human\,/\,aligned-model} pairs.
    Bottom panel: rabbit femur tile with 207 cells, 206 correct
    (99.5\% accuracy).
  }
  \label{fig:tile_comparison}
\end{figure}

\paragraph{Tile-level visual comparison.}
Figure~\ref{fig:tile_comparison} shows a tile-level post-alignment
comparison.
Each cell polygon is overlaid with its human label and aligned model
label (format: \emph{human\,/\,model}).
Green polygons indicate matches; red indicates mismatches.
On this tile (rabbit femur, 207~cells), only 1 cell is mismatched,
yielding 99.5\% accuracy.

\subsection{Analysis of Challenging Cases}
\label{sec:challenges}

The two tissue types with accuracy below 90\%---compact bone
(84.0\%) and skeletal muscle (84.0\%)---share characteristics
that challenge unsupervised clustering:

\begin{itemize}
  \item \textbf{Compact bone} contains very few cells per tile (often
        $<20$), making DBSCAN density estimates unreliable.
  \item \textbf{Skeletal muscle} includes morphologically diverse
  tissue components---individual muscle fibres, nuclei, and other
  connective-tissue elements---that a human annotator can distinguish
  using spatial context.
  The model, operating on cropped single-object appearance without
  spatial awareness, may group visually similar but biologically
  distinct components into the same cluster.
\end{itemize}

These failure modes suggest that tissue-specific parameter tuning
(\eg adjusting $\varepsilon$ or $\mathtt{min\_samples}$) or
incorporating spatial context could improve performance on
challenging tissues.

\section{Discussion}
\label{sec:discussion}

\paragraph{Practical impact.}
The cluster-first paradigm fundamentally changes the annotation
workflow.
Instead of $\mathcal{O}(N)$ labelling effort where $N$ is the number
of cells, the annotator's effort scales with the number of
clusters~$K$, typically $K \ll N$.
On our evaluation slides, $K$ ranged from 1--3 human-defined classes;
even with DBSCAN producing a modestly higher number of clusters, the
annotator's task reduces from reviewing thousands of cells to reviewing
tens of clusters.

\paragraph{Generality.}
Using a single fixed configuration, the pipeline achieves
$\geq 97\%$ accuracy on 10 of 13 tissue types and $\geq 99\%$ on 9 of
13.
This broad generalisation is enabled by two design choices:
(i)~Cellpose-SAM's tissue-agnostic segmentation, and
(ii)~the combination of ResNet-50 feature extraction with UMAP
compression, which produces a compact embedding space where
morphologically similar objects are nearby and dissimilar objects are
well-separated, allowing a single set of DBSCAN parameters to
generalise across tissue types.
Notably, the pipeline is not restricted to individual cells: it
operates on any tissue component detected by Cellpose-SAM, including
nuclei, cell clusters, and other structures.
The clustering step naturally separates these heterogeneous detections
into coherent groups, allowing downstream annotation tools to label or
exclude entire categories at the cluster level rather than per object.

\paragraph{Limitations.}
The evaluation covers 3{,}696 segmented objects across 29 tiles---sufficient
to demonstrate the approach but not exhaustive.
Because Cellpose-SAM detects a variety of cell-like structures, the evaluated
objects include individual cells, nuclei, and potentially other tissue
components; we measure clustering agreement rather than biological
classification correctness.
Segmentation quality (\ie whether boundaries are pixel-accurate) is
outside the scope of this work; we evaluate only whether morphologically
similar objects are grouped together.
Upstream segmentation errors such as under- or over-segmentation may
propagate to clustering by splitting or merging objects, but this effect
is not quantified.
The LLM-based classification stage remains experimental and is not
evaluated in this work.
Performance on tissues such as compact bone or skeletal muscle
may benefit from tissue-specific tuning or more spatial awareness.

\paragraph{Open-source release.}
All pipeline code, the labelling web application, and evaluation
scripts are released under the MIT licence at
\url{https://github.com/OxfordCompetencyCenters/edu06_histology_labelling}
(DOI: \href{https://doi.org/10.5281/zenodo.18927168}{10.5281/zenodo.18927168}).

\section{Conclusion}
\label{sec:conclusion}

We presented a fully automated pipeline for cell segmentation and
morphological clustering in histology whole slide images, together with
a human evaluation framework based on Hungarian-algorithm alignment.
Evaluated on 3{,}696 cells spanning 13 tissue types and three species,
the pipeline achieves 96.8\% cluster--label alignment accuracy with a
single fixed configuration.
The \emph{cluster-first} paradigm reduces annotation effort from
thousands of individual cells to tens of representative clusters,
making large-scale histology annotation practical.
Further exploration of integration of spatial context may improve 
performance on challenging tissues.

\section*{Acknowledgements}

We thank the University of Oxford, AI Competency Center and the
Medical Sciences Division for supporting this
research.

\bibliographystyle{plain}
\bibliography{references}

\begin{thebibliography}{10}

\bibitem{bankhead2017qupath}
Peter Bankhead, Maurice~B Loughrey, Jos{\'e}~A Fern{\'a}ndez, Yvonne
  Dombrowski, Darragh~G McArt, Philip~D Dunne, Stephen McQuaid, Ronan~T Gray,
  Liam~J Murray, Helen~G Coleman, et~al.
\newblock {QuPath}: Open source software for digital pathology image analysis.
\newblock {\em Scientific Reports}, 7(1):16878, 2017.

\bibitem{bradski2000opencv}
Gary Bradski.
\newblock The {OpenCV} library.
\newblock {\em Dr.\ Dobb's Journal of Software Tools}, 2000.

\bibitem{chen2024pathology}
Richard~J Chen, Tong Ding, Ming~Y Lu, Drew F~K Williamson, Guillaume Jaume,
  Andrew~H Song, Bowen Chen, Andrew Zhang, Daniel Shao, Muhammad Shaban, et~al.
\newblock Towards a general-purpose foundation model for computational
  pathology.
\newblock {\em Nature Medicine}, 30(3):850--862, 2024.

\bibitem{deng2009imagenet}
Jia Deng, Wei Dong, Richard Socher, Li-Jia Li, Kai Li, and Li~Fei-Fei.
\newblock {ImageNet}: A large-scale hierarchical image database.
\newblock In {\em IEEE Conference on Computer Vision and Pattern Recognition},
  pages 248--255, 2009.

\bibitem{doron2023unbiased}
Michael Doron, Th{\'e}o Moutakanni, Zitong~S Chen, Nikita Moshkov, Mathilde
  Caron, Hugo Touvron, Piotr Bojanowski, Wolfgang~M Pernice, and Juan~C
  Caicedo.
\newblock Unbiased single-cell morphology with self-supervised vision
  transformers.
\newblock {\em bioRxiv}, 2023.

\bibitem{ester1996dbscan}
Martin Ester, Hans-Peter Kriegel, J{\"o}rg Sander, and Xiaowei Xu.
\newblock A density-based algorithm for discovering clusters in large spatial
  databases with noise.
\newblock In {\em Proceedings of the 2nd International Conference on Knowledge
  Discovery and Data Mining}, pages 226--231, 1996.

\bibitem{farahani2015wsi}
Navid Farahani, Anil~V Parwani, and Liron Pantanowitz.
\newblock Whole slide imaging in pathology: advantages, limitations, and
  emerging perspectives.
\newblock {\em Pathology and Laboratory Medicine International}, 7:23--33,
  2015.

\bibitem{goode2013openslide}
Adam Goode, Benjamin Gilbert, Jan Harkes, Drazen Jukic, and Mahadev
  Satyanarayanan.
\newblock Openslide: A vendor-neutral software foundation for digital
  pathology.
\newblock {\em Journal of Pathology Informatics}, 4(1):27, 2013.

\bibitem{graham2019hovernet}
Simon Graham, Quoc~Dang Vu, Shan E~Ahmed Raza, Ayesha Azam, Yee~Wah Tsang,
  Jin~Tae Kwak, and Nasir Rajpoot.
\newblock Hover-net: Simultaneous segmentation and classification of nuclei in
  multi-tissue histology images.
\newblock {\em Medical Image Analysis}, 58:101563, 2019.

\bibitem{he2016resnet}
Kaiming He, Xiangyu Zhang, Shaoqing Ren, and Jian Sun.
\newblock Deep residual learning for image recognition.
\newblock In {\em Proceedings of the IEEE Conference on Computer Vision and
  Pattern Recognition}, pages 770--778, 2016.

\bibitem{kuhn1955hungarian}
Harold~W Kuhn.
\newblock The {Hungarian} method for the assignment problem.
\newblock {\em Naval Research Logistics Quarterly}, 2(1--2):83--97, 1955.

\bibitem{lu2021clam}
Ming~Y Lu, Drew F~K Williamson, Tiffany~Y Chen, Richard~J Chen, Matteo
  Barbieri, and Faisal Mahmood.
\newblock Data-efficient and weakly supervised computational pathology on
  whole-slide images.
\newblock {\em Nature Biomedical Engineering}, 5(6):555--570, 2021.

\bibitem{mcinnes2018umap}
Leland McInnes, John Healy, and James Melville.
\newblock {UMAP}: Uniform manifold approximation and projection for dimension
  reduction.
\newblock {\em arXiv preprint arXiv:1802.03426}, 2018.

\bibitem{azureml2023}
{Microsoft}.
\newblock Azure {Machine Learning} documentation.
\newblock \url{https://learn.microsoft.com/en-us/azure/machine-learning/},
  2026.
\newblock Accessed: 2026-04-09.

\bibitem{pachitariu2025cellposesam}
Marius Pachitariu, Michael Rariden, and Carsen Stringer.
\newblock Cellpose-{SAM}: superhuman generalization for cellular segmentation.
\newblock {\em bioRxiv}, 2025.

\bibitem{pantanowitz2011wsi}
Liron Pantanowitz, Paul~N Valenstein, Andrew~J Evans, Keith~J Kaplan, John~D
  Pfeifer, David~C Wilbur, Laura~C Collins, and Terence~J Colgan.
\newblock Review of the current state of whole slide imaging in pathology.
\newblock {\em Journal of Pathology Informatics}, 2(1):36, 2011.

\bibitem{rapids2019}
{RAPIDS Development Team}.
\newblock {RAPIDS}: Open {GPU} data science.
\newblock \url{https://rapids.ai}, 2026.
\newblock Accessed: 2026-04-09.

\bibitem{ronneberger2015unet}
Olaf Ronneberger, Philipp Fischer, and Thomas Brox.
\newblock U-net: Convolutional networks for biomedical image segmentation.
\newblock In {\em Medical Image Computing and Computer Assisted Intervention --
  MICCAI 2015}, pages 234--241. Springer, 2015.

\bibitem{schmidt2018stardist}
Uwe Schmidt, Martin Weigert, Coleman Broaddus, and Gene Myers.
\newblock Cell detection with star-convex polygons.
\newblock In {\em Medical Image Computing and Computer Assisted Intervention --
  MICCAI 2018}, pages 265--273. Springer, 2018.

\bibitem{srinidhi2021deep}
Chetan~L Srinidhi, Ozan Ciga, and Anne~L Martel.
\newblock Deep neural network models for computational histopathology: A
  survey.
\newblock {\em Medical Image Analysis}, 67:101813, 2021.

\bibitem{stringer2021cellpose}
Carsen Stringer, Tim Wang, Michalis Michaelos, and Marius Pachitariu.
\newblock Cellpose: a generalist algorithm for cellular segmentation.
\newblock {\em Nature Methods}, 18(1):100--106, 2021.

\end{thebibliography}

\end{document}